\documentclass[conference]{IEEEtran}
\usepackage{cite}
\usepackage[export]{adjustbox}
\usepackage{balance}

\ifCLASSINFOpdf
\else
\fi
\usepackage{amsmath}
\usepackage{amsfonts}
\usepackage{eucal}
\usepackage{algorithmic}

\usepackage{array}
\begin{document}
%
\title{Relay Assisted Device-to-Device Communication: Approaches and Issues}

\author{\IEEEauthorblockN{Uyoata Uyoata\IEEEauthorrefmark{1} and
Mqhele Dlodlo\IEEEauthorrefmark{2}}
\IEEEauthorblockA{Department of Electrical Engineering, University of Cape Town, Cape Town, South Africa\\ 
Email: \IEEEauthorrefmark{1} uytuyo001@myuct.ac.za,
\IEEEauthorrefmark{2} Mqhele.dlodlo@uct.ac.za}}


%


\maketitle

\begin{abstract}
Enabling technologies for 5G and future wireless communication have attracted the interest of industry and research communities. One of such technologies is Device-to-Device (D2D) communication which exploits user proximity to offer spectral efficiency, energy efficiency and increased throughput. Data offloading, public safety communication, context aware communication and content sharing are some of the use cases for D2D communication. D2D communication can be direct or through a relay depending on the nature of the channel in between the D2D devices. Apart from the problem of interference, a key challenge of relay aided D2D communication is appropriately assigning relays to a D2D pair while maintaining the QoS requirement of the cellular users. In this article, relay assisted D2D communication is reviewed and research issues are highlighted. We also propose matching theory with incomplete information for relay allocation considering uncertainties which the mobility of the relay introduces to the set up. 
\end{abstract}


\begin{IEEEkeywords}
Device-to-Device communication, relay allocation, channel uncertainty, mobile relaying.
\end{IEEEkeywords}

%
\IEEEpeerreviewmaketitle

\section{Introduction}
Wireless communication has evolved in to a ubiquitous technology with widespread applications. It is estimated that future wireless communication technologies will need to cater for much higher speed and reduced latency than the present 4G networks. To enable the evolution towards 5G networks and beyond, D2D communication within cellular networks has been proposed as an enabling technology and has been included in third generation partnership project (3GPP)’s Proximity Services (ProSe) standard \cite{6807945}. D2D communication could be deployed as an overlay \cite{5074679} to an existing cellular network where it uses orthogonal channels or as an underlay in which available resources are shared \cite{5350367}. The former reduces interference while the later makes for better spectral efficiency. The gains associated with D2D communication include hop gain, reuse gain, and proximity gain \cite{6163598}. While direct communication between devices is not new, D2D communication employs the already available network and so overcomes the manual pairing and access point definition associated with unlicensed spectrum communication. Studies in D2D communication have considered the feasibility of D2D underlaying cellular networks\cite{5073734}, resource allocation, spectrum sharing \cite{5506183} and coverage extension \cite{7227122}. Cooperative communication has also been considered in \cite{6831680} \cite{6952682}. D2D communication can be direct between two mobile devices or relay assisted in which either a fixed low power relay is employed to forward signals or a mobile relay is used. The need to introduce relay to D2D communication arises when the distance between the nodes is too far for direct communication or when the channel is severely impaired. Instances where relay assisted D2D communication can be used include: coverage extension and in content sharing networks.


For cellular communication, 3GPP has standardized radio relay technologies specifying Layer 1 Relay, Layer 2 Relay and Layer 3 Relay. The standardization of D2D communication is still in the works and associated relay techniques have not yet been standardized. A cell with multiple technologies is shown in Fig.1. This article reviews the approaches to relay aided D2D communication with a view of highlighting important research issues. 
This paper is structured as follows, Section II describes LTE-A relay technologies, an overview of D2D communication is given in section III,  whereas resource allocation and performance analysis in relay aided D2D communication are considered in section IV.Section V outlays relay selection approaches in D2D communication while the application scenarios are presented in section VI. Research issues are discussed in section VII and the effect of channel uncertainties due to mobility is studied in section VIII. The paper is concluded in section IX.


%

\section{LTE-A Relay Technology}
Relay technology has been standardized by 3GPP for LTE-A with the aim of improving cell edge throughput and improving cell coverage. A relay provides access to UEs in the downlink and a wireless backhaul to the base station in the uplink. Radio relay technologies can be broadly classified as Layer 1 Relay, Layer 2 Relay and Layer 3 Relay.  A layer 1 relay is simply a repeater station which amplifies its received radio signal and forwards it to the destination.  Layer 2 relays amplifies received signals only after successful decoding/encoding and demodulation/modulation.  Layer 3 relay is similar to Layer 2 relay but having radio protocols similar to an LTE base station.  \cite{iwamura2010relay}. A summary of the features of the relay technology types is given in table 1. To adapt LTE-A relay technologies to D2D communication, consideration has to be made of whether the device acting as a relay is a fixed infrastructure of the type in Table 1 or a UE.

\begin{table*}[htb]
\caption{LTE-A RELAY TECHNOLOGY}
\centering
\begin{tabular}{|c|c|c|c|}
\hline
Relay Technology & Features & Pros & Cons\\
\hline
Layer 1 Relay & Amplify and Forward  & Simplicity of implementation  & Noise and interference  amplification\\
           &         &  Minimal processing delay & \\
\hline
Layer 2 Relay & Decode and Forward  & Noise and interference removal & Processing delay due to decoding and encoding\\
\hline
Layer 3 Relay & Decode and Forward  & Noise and interference removal & Additional processing delay due to: \\
&  Modulation/Demodulation & & decoding and encoding and user data processing \\
&  User data regeneration processing & & \\
\hline
\end{tabular}
\end{table*}

\begin{figure}[!t]
\includegraphics[width=3.0in]{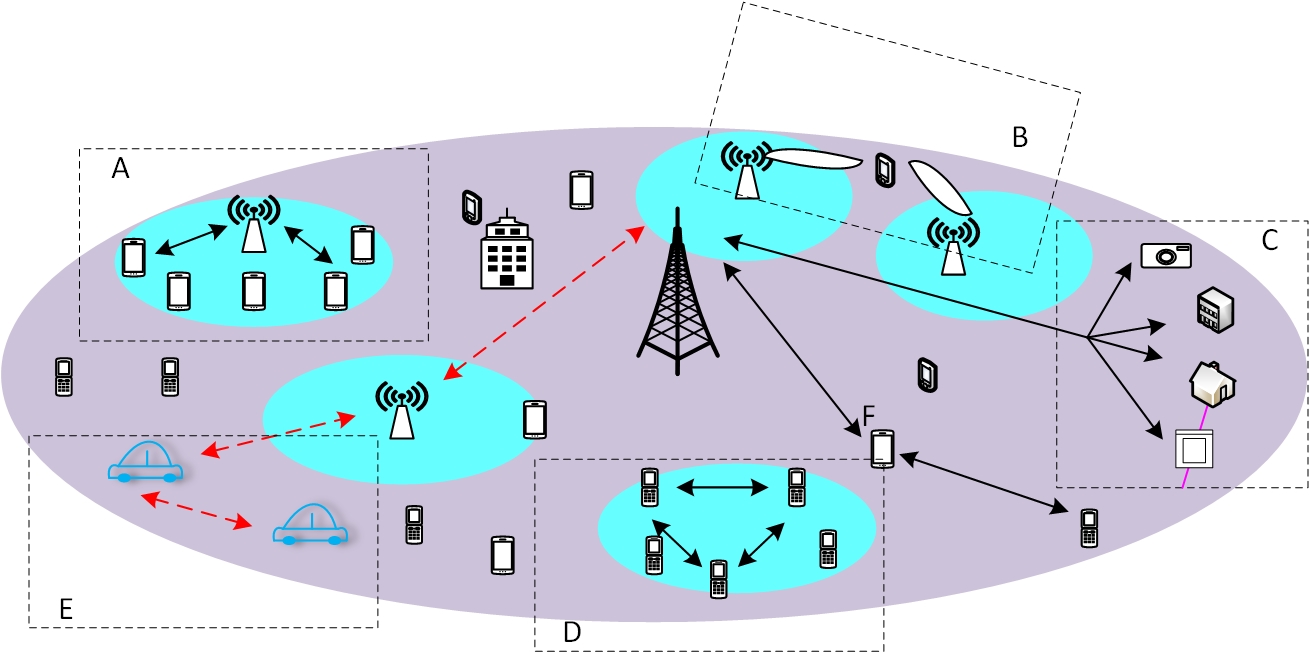}
\caption{Multiple technologies overlaid in a macro cell; A: Fixed relay assisted D2D communication; B:mmW communication;C:Massive machine type communication; D:Mobile relay assisted D2D/D2D Mesh;E: short range inter-vehicular communication; F: Mobile relaying }
\end{figure}

\section{D2D Communication}
A technology envisioned for future wireless communication is device to device (D2D) communication. Introduced in \cite{832516} as a means of allowing nearby devices communicate through multi-hops instead of a single hop via the base station D2D communication is now a technology item for LTE-Advanced. D2D communication has received attention in the academia, industry and has seen standardization efforts by the 3GPP. An actual implementation of a cellular D2D communication was FlashLinQ; a creation of Qualcomm that exploits the orthogonality in OFDM to provide peer-to-peer communication among devices in close proximity \cite{6544312}. Primarily the instances where D2D communication can be strongly advocated for is in public safety communication when the cellular infrastructure is damaged, content sharing in close proximity (like multimedia in dens area), gaming among others. It allows for ad-hoc networks to be formed between user devices with little or no coordination from the macro cell or operator base station. D2D communication extends the fixed relaying which is already a part of the LTE-Advanced standard \cite{6146495}  to mobile relaying. Mesh networks in which device communicated directly are not new; Bluetooth, WLAN,HiperLAN2, TETRA, Infra-red, NFC are technologies which have allowed such communication in the unlicensed band although these largely are deployed in an unplanned nature. The 3GPP in an effort to integrate D2D communication in the licensed band into its standard has created study groups to research in this area \cite{___}. Deployment of D2D communication in the same band as the macro-cell base station (and by extension other co-channel small cells), interference resource allocation, security valid concerns. Although deployed within the cellular network, D2D communication is not completely similar to UE to eNodeB(or BS) links especially with respect to the channel model. In D2D communication, both the transmitter and receiver are mobile in contrast to the stationary cellular BS thus affecting the nature of the channel fading and the temporal correlation of shadowing. Furthermore the transmitter and receiver heights are lower with respect to base station heights and the distance between the devices are smaller considering envisioned dense deployment \cite{6807945}. Resource management wise, managing D2D communication depends on whether the devices are within the coverage of a network or not. For the case where the D2D UEs are within the coverage of an eNodeB/BS, the base station provides management and control functions whereas outside the coverage areas, the devices can act in ad hoc mode or a device can assume the position of a cluster head although the later drains the battery of the device and so there could be incentives to encourage such. But from a public safety network perspective, an incentive may not be considered. D2D communication has been shown to offer various gains to the overall network performance by way of, increased user throughput \cite{6807945}, \cite{6787081}, \cite{5350367}, traffic offloading \cite{5350367}, \cite{6692723}, spectrum utilization \cite{5074679}, \cite{5350367}. 

Although D2D communication targets close proximity devices, in a situation where the channel between the devices is impaired, relaying can be used to assist in forwarding the intended signals.  Relays can be fixed or mobile. When relays are in motion at high speeds, obtaining accurate channel state information becomes difficult due to the channel variation rate. In the literature, the approaches in analysing relay aided D2D communication can be categorised into: resource allocation, relay selection and applications.
%

\section{Resource Allocation and Performance Analysis Approaches}
The envisaged dense and heterogeneous nature of future wireless networks have motivated wide research into resource allocation and interference management. For D2D communication, algorithms for mode selection, resource allocation, and power control have also been proposed. While most resource allocation algorithms have been centralised, distriibuted techniques have been developed for the seemingly ad-hoc nature of D2D communication. In \cite{6848847}, resource allocation for relay based D2D communication is considered in which devices pass messages D2D nodes exchange messages with the relay to determine suitability of transmission. The LTE-A L3 relay is used and sum rate is the metric for performance measure although relay selection is not considered. Still using LTE L3 Relay\cite{6825052}, numerical optimization is used for a centralised resource allocation in relay aided D2D communication. The optimization problem is formulated as a mixed integer non-linear problem for which relaxations are made for ease of analysis. Perfect channel estimations are assumed and a distance threshold is obtained beyond which using a relay does not offer the target gains. Still using the LTE-A L3 Relay as a platform the authors in \cite{6825052} propose an algorithm robust against channel uncertainties which is modelled using a bounded uncertainty set with known probability \cite{6775376}. The assumed sources of uncertainty are  channel gains and interference at receiving ends.  Although assuming a perfect channel can ease analysis, such analysis serve as benchmarks to those that consider imperfect channel conditions. Game theoretical approaches have been used for resource allocation in wireless communication. In \cite{7182781} a promising approach namely matching theory is used for distributed resource allocation. The channel uncertainty is modelled as in \cite{6775376} and LTE L3 Relay is also used. 
While resource allocation has attracted the interest of various researchers, other articles have focused on theoretical analysis and performance evaluation of relay based D2D communication. Using Monte Carlo simulations, Kiran Vanganuru et al \cite{6415659}  demonstrate the potential of mobile relaying in improving the capacity and coverage of a network. Decode and forward (DF) cooperative protocol is used at the mobile relay and the eNB centrally controls the assigning of relays. Considering multi-hop through UEs, Donghoon Lee et al \cite{6400975} performed outage probability and ergodic capacity analysis for decode and forward D2D communication showing the gains thereof. The analysis assumes a single D2D pair and a single interfering cellular UE.  Stochastic analysis is the approach employed by Akeam Al-Hourani et al \cite{7386667} with the aim of proposing the optimum relay region within which energy is saved. Departing from other relay region formulation results, it proposes an elliptical relay region. Furthermore, a distance threshold is presented that provides the boundary beyond which relay assisted communication is not preferable.  
\section{Relay Selection In Relay Aided D2D Communication}
Most of the literature considering resource allocation in relay aided D2D communication have either focused on a single fixed relay or multiple relays whose presence are either viewed as sources of interference or as part of a multi-hop chain. For efficient resource utilization, selecting a single relay out of a cluster of relays is also of importance. In \cite{6362495} a non-centralised approach to relay selection is proposed in which a mobile relay is selected such that the interference from d2d communication to cellular UEs are minimized. The relay selection is timer based such that the relay experiencing least interference has the least time to forward signals from a D2D transmitter. Perfect channel estimation is assumed and multiple cellular links with a single D2D link is considered. Extending the work in \cite{6362495}, Chen Zhengwen et al \cite{6946156} formulated relay selection as an optimization problem in which the capacity of the relay-D2D receiver link is optimized while ensuring the QoS requirement of cellular UEs. The relay selection is done jointly with resource allocation to limit interference to cellular UEs. The paper assumes cognitive sensing at the relays and D2D nodes. Weinchen Xia et al \cite{6617508} proposed the use of both CSI and the distance between nodes as criteria for relay selection depending on whether D2D communication is indoor or outdoor respectively. Firstly a cluster of candidate relays is created using the distance/CSI criteria, then the specific relay is selected based on a defined channel efficiency. The metrics of performance are error probability and outage probability. Resource allocation is not the focus of the amplified and forward based work and perfect channel knowledge is assumed. In Boijang Ma et al \cite{7511462}, the problem of relay allocation is formulated as a matching problem between d2d pairs and candidate fixed relays. Bipartite graph is used and the aim is to minimize the power consumption of D2D communication. Yicha Chen et al \cite{7047325}  formulated the relay selection problem as bipartite graph for which an auction algorithm was proposed in which D2D pairs submitted bids for the candidate relay. The proposed algorithm shows increased throughput as against centralised relay allocation algorithm although complexity analysis was not performed. The issue of relay incentives is addressed in \cite{7181721} where UEs through a learning algorithm determine the relay policy which utilizes their utility. The literature uses token passing between UEs and the proposed learning algorithm are tested in scenarios of various mobility. Perfect channel knowledge is assumed and the decision to relay is based on optimizing the relay utility under amplify and forward cooperative protocol. 

\section{Relay Assisted D2D Communication Application Scenarios }
The use cases for relay assisted D2D communication include public safety communication and popular content distribution. Hao Xu \cite{7420936} investigated the use of relay assisted D2D communication in distributing popular video contents within clusters of devices. Inter-cluster interference is considered and a UE relay is introduced when the distance between the D2D nodes is exceeds a threshold. The channel is assumed perfect and the effect of group mobility is not considered. 
Although dedicated public safety communication systems such as Terrestrial Trunked Radio (TETRA) are already available, their speeds do not suit emerging public safety scenarios. G. Fodor et al \cite{6985517} proposed an architecture for d2d communication for public safety and disaster scenarios application. They proposed a cluster based approach where a cluster assumes the functionalities similar to base station functionalities for nodes that are out of coverage. Such roles include resource management and synchronisation. In situations where the elected cluster head is out of coverage, a UE within coverage assumes the roles. The architecture proposed serves full, partial and out of coverage scenarios. The application of mobile devices specifically smartphones in multi-hop relaying in disaster areas is proposed in \cite{6807947} where adaptive algorithm which switches between two routing techniques namely mobile ad-hoc network (MANET) and disruption tolerant network(DTN) depending on the distribution of the mobile devices is presented. Message relaying through a developed mobile application was tested between devices and an unmanned aircraft system was used to aid relay to a distant location. Mobile relaying can present security risks.  In \cite{7247255} beamforming at fixed relay is proposed as a technique to counter attempts by eavesdroppers to intercept information between a D2D pair. Physical layer network coding is employed at the relay node. Moreover  channel uncertainty is considered between the eavesdropper and the entities involved in D2D communication although perfect channel estimation is assumed between the D2D entities. 

\begin{table*}[htb]
\caption{Key Related Literature}
\centering

\begin{tabular}{|c|c|c|c|c|c|c|}
\hline
Paper & Approach & Solution Proposed & Uncertainty &Optimality &coordination& Relay Mobility\\
\hline
\cite{6848847}& Resource allocation  & message passing  & No &Asymptotically optimal& Distributed& Fixed\\
\hline
\cite{6400975} & Theoretical evaluation and  &  Not applicable (NA) & No & NA & Centralised & Mobile \\
&  performance analysis &   &  &  & & \\     
\hline
\cite{6415659} & \checkmark & \checkmark  & \checkmark & \checkmark &Centralised & Fixed\\
\hline
\cite{6825052}  & Resource allocation & Numerical optimization using & \checkmark & Asymptotically optimal & Centralised & Fixed\\

& & Karush-Kuhn-Tucker & &  & & \\
\hline
\cite{7182781}& \checkmark & Matching Theory & Yes & Near Optimal& Distributed & Fixed \\
\hline
\cite{6775376}& \checkmark & Numerical optimization using & \checkmark &\checkmark & \checkmark & Mobile \\

& & Karush-Kuhn-Tucker & & &  &  \\
\hline
\cite{6946156}& Resource allocation and  & Numerical analysis & No & Optimal & Distributed & \checkmark \\
&  Relay selection & &  &  & &  \\
\hline
\cite{7181721}& Relay selection & Numerical analysis& No & NA & Distributed & \checkmark\\
\hline
\cite{6617508} & \checkmark & Performance analysis & \checkmark & \checkmark & Distributed & \checkmark\\
\hline
\cite{7247255} & Interference Management & MIMO beamforming & Yes & & Centralised  & Fixed \\
\hline
\cite{7047325} & Relay allocation & auction based bipartite matching & No & Optimal & Centralised & Mobile\\
\hline
\cite{7511462} &\checkmark & Bipartite matching & \checkmark & Suboptimal & Centralised & Fixed \\

 &\ & Hungarian algorithm & &  & &  \\
\hline

\end{tabular}
\end{table*} 

\section{Research Issues}
Given the forgoing, research in relay based D2D communication is thriving although there are research issues that need be highlighted. While most research works in this area have focused on fixed relays a few have studied the use of mobile phones as the relay. Authors studying the use of UEs as relays generally assume that the UEs are within efficient relaying distance and have not analysed the efficient relay region for mobile relays.  In studying mobile relaying most authors have assumed perfect channel estimation at the relay UE and the transmitting UE. This ignores the uncertainty which channel ageing due to mobility could introduce into the channel formulation. Where uncertainty is considered in the literature, uncertainty is introduced in the interference channel while the channel between a D2D transmitter and relay is assumed perfect. For direct communication such relaxation could hold considering the inter-distance, but for relay based communication, the channel uncertainties may occur. Some works have studied D2D communication in the light of 3GPP LTE/LTE-A standards. They have done so by using already standardized techniques such the 3GPP power control and Layer 3 Relay. Doing so can help fast track standardization procedure for relay aided D2D communication. While research into ensuring secured direct communication between D2D nodes is gaining steam, few works have considered security in relay assisted D2D communication. An overview of security in D2D communication with proposed solutions is given in \cite{6807948}. Moreover while resource allocation, relay selection and performance analyses have been the focus of most papers, energy efficiency of relay assisted D2D communication has not been so much focused on. Furthermore most works assume that the relays and the D2D devices have single antennas. A few works have considered designing beamforming precoders for MIMO D2D communication to avoid intra-cell interference. Without loss of generality, a feature of static nodes in wireless communication is perfect channel estimation which allows for coherent detection. Mobility of users in a wireless network introduces time variation leading to errors in the channel estimation. When channel uncertainty introduced by device mobility is considered in D2D communication  and a relay protocol such as the two way amplify and forward is used for a time division duplex system, the self-interference at the UEs are not completely cancelled due to the channel uncertainties. This makes the analysis less trivial. In the next section we show the effect of such channel uncertainty in relation to the outage probability. A summary of the research works on relay assisted D2D communication is provided in Table II.

\section{Relay Assisted D2D Communication And Channel Uncertainties}
To show the effect of channel variation, a coverage extension scenario is considered in which there is a D2D UE $s_1$ at the cell edge as in Fig.2. By using direct communication between $s_1$ and an appropriately selected idle devices $d_1$ in the range of $s_1$, the coverage of BS can be extended to $s_1$. Fig. 2 shows the system model. Time division duplex with two time slots is considered. In two way relaying, the relay broadcasts in the second time slot a scaled combination of the signals it received from two transmitters in the first time slot such that at the two transmitters cum receivers, the self-interference can be subtracted or cancelled and the target signal extracted. 
The channels between $s_1$ and the $d_1$ is given by h and between $d_1$ and the BS/eNB is given by g. 

\begin{figure}[!t]
\includegraphics[width=3.0in]{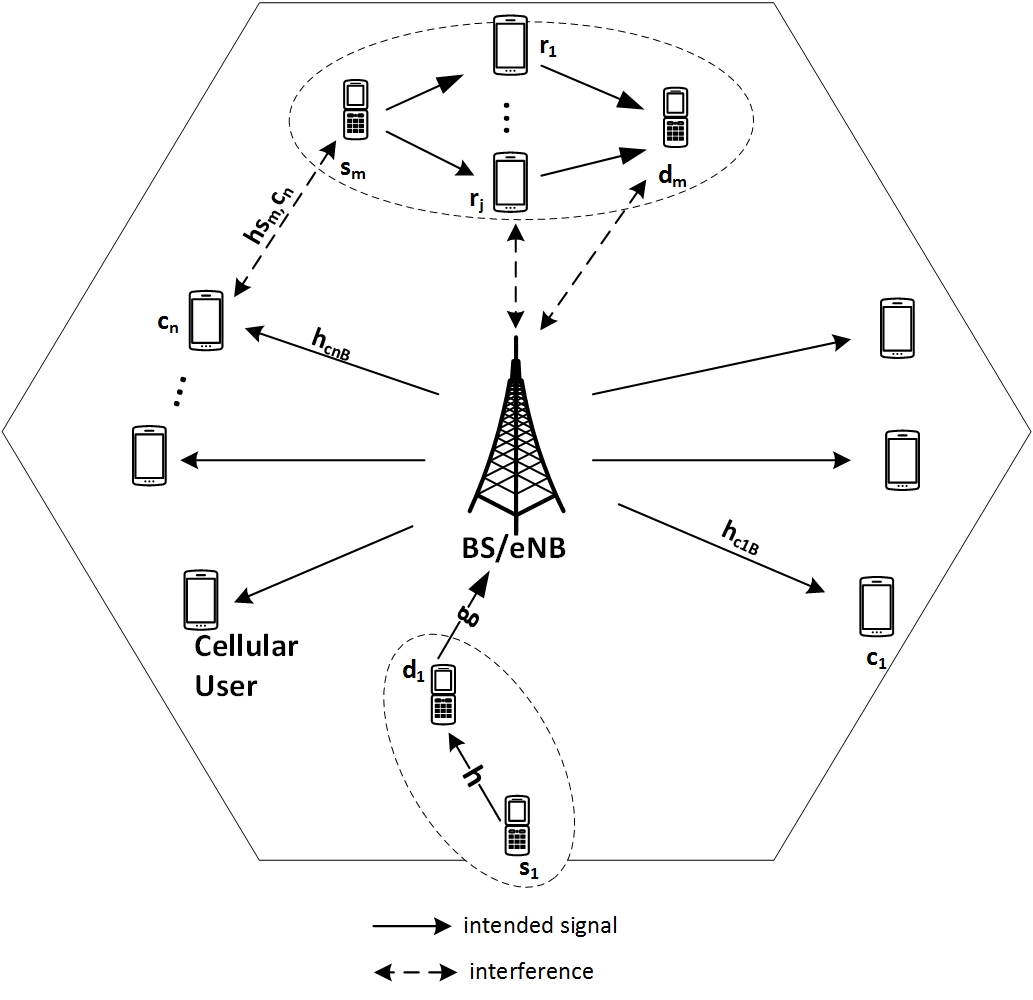}
\caption{Relay Assisted D2D Communication System Model}
\end{figure}

\subsection{Problem Formulation}
For a static or fixed relay, h and g can be modelled as circularly complex with zero mean and a variance value but for this case in which $d_1$ is mobile , the channels must capture the uncertainty which relay mobility introduces. When the relay is in motion, the channel estimation is not perfect due to the time variation of the channel. Hence the channels h and g can be expressed as \cite{1683591}

%
%



\begin{equation}
h(n)= a_1 h(n-1)+\Delta h(n) \sqrt{(1-a_{1}^2 }
\end{equation}
\begin{equation}
g(n)= a_2 g(n-1)+\Delta g(n) \sqrt{(1-a_{2}^2)}
\end{equation}
Where $a_i$, i=1,2 indicates the channel variation rate related to the Doppler shift by the zeroth order Bessel function of the first kind (i.e. $a[n]= J_{0}(2\pi f_{D}T[n]))$, while $\Delta h(n)$ and $\Delta g(n)$ are the time varying component of the channel h and g and are independent and identically distributed with distribution $\CMcal{CN}(0,\sigma_{i}^2), i=1,2$ \cite{1512123}. Estimation (that is the knowledge the receiver has) of the channels are given by $\widehat{h}$ and $\widehat{g}$ are complex random Gaussian processes described as, $\widehat{h}\sim\CMcal{CN}(a_{1} h_{o}(1-a_{1}^2)\sigma_{1}^2), \widehat{g}\sim \CMcal{CN}(a_{2} g_{o}(1 - a_{2}^2)\sigma_{2}^2)$. Without loss of generality it can be inferred that at an instant $t_{o}$ where there is no channel variation or rather at the start of transmission, $\widehat{h}=h_o$ and $\widehat{g} = g_{o}$. 

Assuming for sake of tractability of analysis that the interference at $s_1$ and BS are ignored, the measured SINRs at the $s_1$ and BS can be expressed as:
%

 \begin{align}
\gamma_1=\beta^2 \alpha^2_1P_2|h_0|^2|g_0|^2  \div \bigg(2\beta^2 \alpha^2_1P_1|h_0|^2(1-\alpha^2_1)+ ... \notag \\
\beta^2 \alpha^2_1|h_o|^2\sigma_r^2+\beta^2 \alpha^2_1P_2|g_0|^2(1-\alpha_1^2)+\sigma_1^2 \bigg)
\end{align}

\begin{equation}
\gamma_2=\dfrac{\beta^2 \alpha^2_1P_1|h_0|^2|g_0|^2}{\beta^2 \alpha^2_1P_1|g_0|^2(1-\alpha^2_1)+ \beta^2P_1|g_o|^2+\sigma_2^2}
\end{equation} 

Where $P_i$ is the transmit power of node $i$, $\beta$ is the amplification factor at $s_1$ associated with amplify and forward relaying, $\sigma_n$ is the noise variance at node n.
%
Denote data rates at  $s_1$ and BS as $R_1$ and $R_1$ respectively   Therefore for the data rate at  the outage probability $P_{out}$ given a target/threshold data rate of $R_{th}=5bits/s/hz$ can be expressed as;

\begin{equation}
P_{out}=P_r(min(R_1,R_2)<R_{th})
\end{equation} 

The probability that the cell edge user $s_1$ is in outage when a helper node with which it forms a D2D pair is in mobility is investigated in this section. Firstly the impact of such mobility is demonstrated for selected variation rate of the channel $(a_1 = 0.998, 0.899, 0.799)$. Fig 3 shows a sharp increase in outage probability as the values of $a_1$ approaches 0 which is not counter intuitive.  Fig. 3 demonstrates the effect of time variation in mobile relaying. Considering that future cellular network will be dense and the mobility of devices are largely independent, the variation rate of the channels between a D2D transmitter and possible receivers (in this case potential mobile relay) vary. The effect of relay mobility points to the need to factor in the uncertainty which mobility introduces into D2D communication channel in analysis and so develop robust algorithms for relay allocation. An approach that can be viable is matching theory with incomplete information. 
\begin{figure}[!t]
\includegraphics[width=3.0in]{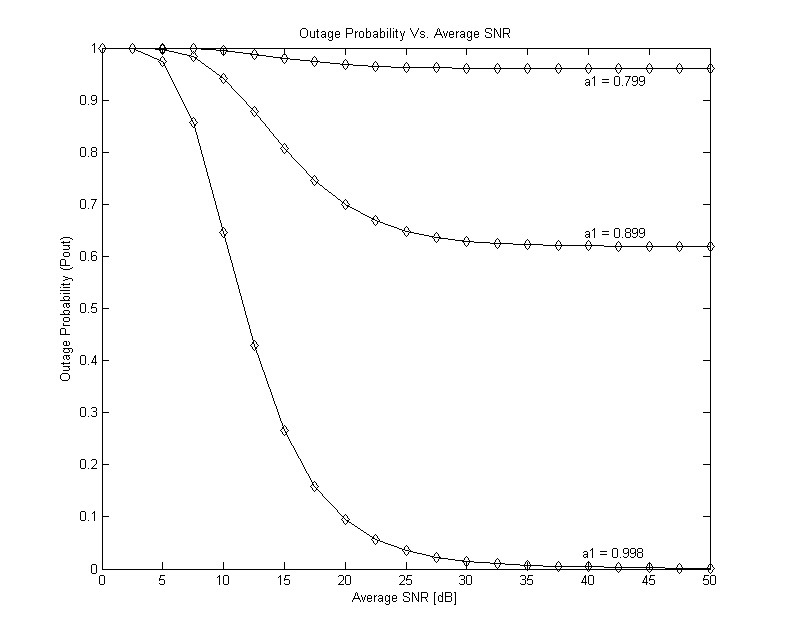}
\caption{Probability of outage vs. SNR}
\end{figure}

\section{Conclusion}
In this work, a review of key literature on relay assisted has been provided highlighting addressed issues and open areas to further explore. We have made a case for the channel uncertainties in D2D communication resulting from device mobility and suggested matching theory with incomplete information as a viable approach to such uncertainties. In a sequel, matching with incomplete information will be employed to achieve desired relay allocation to the D2D links.


%



\bibliographystyle{IEEEtran}
\bibliography{Uyoata}
%


\end{document}